\documentclass[12pt,letterpaper]{article}

\usepackage[margin=1in]{geometry}
\usepackage{color}
\usepackage{comment}
\usepackage{graphicx}
\usepackage[utf8]{inputenc}
\usepackage{aas_macros}
\usepackage{subfigure}
\usepackage{xcolor}
\usepackage{soul}
\usepackage{graphicx}
\usepackage{floatrow}
\usepackage{amsmath}
\usepackage[T1]{fontenc} 
\usepackage{ amssymb }
\usepackage{xspace}
\usepackage{hyperref}
\usepackage{ragged2e}
\usepackage{sidecap} 
\usepackage{ulem}
\usepackage[maxbibnames=3,natbib=true]{biblatex}
\appto{\bibsetup}{\raggedright}

\def\m87{M87\xspace}                    
\def\uas{$\mu$as\xspace}                
\def\msun{M$_{\odot}$\xspace}           
\def\origins{OST\xspace}                

\begin{document}


\raggedright
\huge
Astro2020 APC White Paper \linebreak

\textbf{Extremely long baseline interferometry with Origins Space Telescope}\linebreak
\normalsize
\justify

\vspace{4mm}
\textsc{Dominic W. Pesce}$^{\dag}{}^{1,2}$, \textsc{Kari Haworth}$^{1}$, \textsc{Gary J. Melnick}$^{1}$, \textsc{Lindy Blackburn}$^{1,2}$, \textsc{Maciek Wielgus}$^{1,2}$,  \textsc{Michael D. Johnson}$^{1,2}$, \textsc{Alexander Raymond}$^{1,2}$, \textsc{Jonathan Weintroub}$^{1}$, \textsc{Daniel C. M. Palumbo}$^{1,2}$, \textsc{Sheperd S. Doeleman}$^{1,2}$, \textsc{David J. James}$^{1,2}$\\[2mm]

\noindent{}\textbf{Abstract:} Operating $1.5 \times 10^6$\,km from Earth at the Sun-Earth L2 Lagrange point, the Origins Space Telescope equipped with a slightly modified version of its HERO heterodyne instrument could function as a uniquely valuable node in a VLBI network.  The unprecedented angular resolution resulting from the combination of Origins with existing ground-based millimeter/submillimeter telescope arrays would increase the number of spatially resolvable black holes by a factor of $10^6$, permit the study of these black holes across all of cosmic history, and enable new tests of general relativity by unveiling the photon ring substructure in the nearest black holes.

\vfill

\footnotesize
\noindent $^{1}$ Center for Astrophysics $\vert$ Harvard \& Smithsonian, 60 Garden Street, Cambridge, MA 02138, USA \\
$^{2}$ Black Hole Initiative at Harvard University, 20 Garden Street, Cambridge, MA 02138, USA

\noindent \line(1,0){250}\\
\textnormal{$^{\dag}$Corresponding author: \href{mailto:dpesce@cfa.harvard.edu}{dpesce@cfa.harvard.edu}}

\normalsize

\newpage

\section{Introduction}

A leap forward in our understanding of black holes came earlier this year, when the Event Horizon Telescope (EHT) collaboration revealed the first horizon-scale image of the active galactic nucleus of \m87 \cite{PaperI}.  This feat was accomplished using a ground-based very long baseline interferometry (VLBI) network with baselines extending across the planet, reaching the highest angular resolution currently achievable from the surface of the Earth. We could soon have an opportunity to make another leap by utilizing the extremely long baselines between Earth and the Origins Space Telescope (\origins).  The corresponding ${\sim}120{\times}$ improvement in angular resolution would increase the expected number of spatially resolvable black hole shadows from ${\sim}1$ to ${\gtrsim}10^6$, enabling new studies of black hole demographics across cosmic history.  For the closest supermassive black holes -- such as the one in \m87 -- the unprecedented angular resolution would yield access to photon ring substructure, providing a new tool for making precise black hole spin measurements and testing the validity of general relativity (GR).

\origins is one of four NASA-funded Flagship mission concepts prepared for the 2020 Decadal Review. \origins would incorporate a 5.9-meter diameter on-axis telescope operating between 2.8 and 588 microns, and it would observe from the Sun-Earth L2 Lagrange point 1.5 million kilometers from Earth.  The primary mirror has a total wavefront error of less than 2.5 microns, making it a superb antenna at (sub)millimeter wavelengths.
 
As proposed to the Decadal Review, \origins is designed with sufficient mass, power, and volume margins to accommodate two additional instruments beyond the baseline complement should the Decadal Review decide their inclusion is desired.  One of these instruments, the HEterodyne Receiver for \origins (HERO, \cite{Hero2018}), has been studied extensively by many of the same people that built the Heterodyne Instrument for the Far-Infrared (HIFI) flown successfully aboard the Herschel Space Observatory.  HERO is presently designed to operate in four bands that span the range between 486 GHz and 2700 GHz, and it would permit both dual-polarization and dual-frequency observations. 

Expanding the HERO instrument to be an interferometric station will require several modest technology enhancements. Top-level discussions have not revealed any fundamental obstacles to these added requirements, which are presented here.

\section{Science case}

Dramatically improving the angular resolution of (sub)mm VLBI is an exciting but daunting prospect that requires either increasing the observing frequency, extending the baseline lengths, or both.  Not many sites on Earth offer atmospheric conditions that are good enough for observations at (sub)mm wavelengths to be routinely viable, and the current longest baselines are already nearly equal to one Earth diameter. Sizable angular resolution improvements will thus inevitably require stations in space.  The L2 orbit planned for \origins provides a unique opportunity for extending VLBI to extremely long baselines by observing in tandem with sensitive ground-based stations (such as ALMA, LMT, NOEMA, GBT, or ngVLA). Such observations would achieve an unprecedented angular resolution.  The EHT is currently the highest-frequency ground-based VLBI network, operating at a frequency of 230 GHz and attaining a resolution of ${\sim}20$\,\uas; by comparison, an Earth-L2 baseline would have typical fringe spacings well under a microarcsecond at observing frequencies of 86, 230, 345, and 690 GHz.

\subsection{SMBH demographics and cosmic evolution}

For a black hole viewed by an observer at infinity, the locus of event horizon-grazing photon trajectories forms a nearly circular\footnote{For a non-spinning black hole the locus is perfectly circular with a radius equal to $\sqrt{27}$ times the Schwarzschild radius.  For spinning black holes that aren't observed pole-on, the locus appears flattened on one side and the radius changes by ${\pm}{\sim}4$\%.} closed curve on the sky \citep{Bardeen_1973}.  This boundary defines the inner edge of the ``photon ring,'' and for astrophysical black holes emission from the black hole ``shadow'' region interior to the photon ring is expected to be substantially depressed (see \citealt{PaperI} and references therein).

Given a uniform distribution of supermassive black holes (SMBHs) in flat space, we expect the number of sources $N$ with spatially resolved black hole shadows to increase as the cube of the maximum baseline length.  Current Earth-based arrays, such as the EHT, are able to spatially resolve black hole shadows for $N \approx 1$ source \citep{PaperI, PaperIII, PaperIV}. \textbf{Extending a baseline from Earth to L2, at a distance of $\boldsymbol{\sim}$120 Earth diameters, would increase the expected number of spatially resolvable black hole shadows\footnote{More precisely, we are considering a~subset of SMBHs for which the shadow could be seen through the surrounding material, which in general depends on the details of the particular accretion flow.} from $\mathbf{\textit{N} \boldsymbol{\approx} 1}$ to $\mathbf{\textit{N} \boldsymbol{\approx} 120^3 \boldsymbol{>} 10^6}$.}  Each black hole with a resolved shadow would have a corresponding black hole mass estimate (or, more specifically, an estimate of the mass-to-distance ratio $M/D$), enabling studies of SMBH mass demographics with access to an unprecedented statistical sample.

\begin{figure}[t]
\centering
\includegraphics[width=0.49\textwidth]{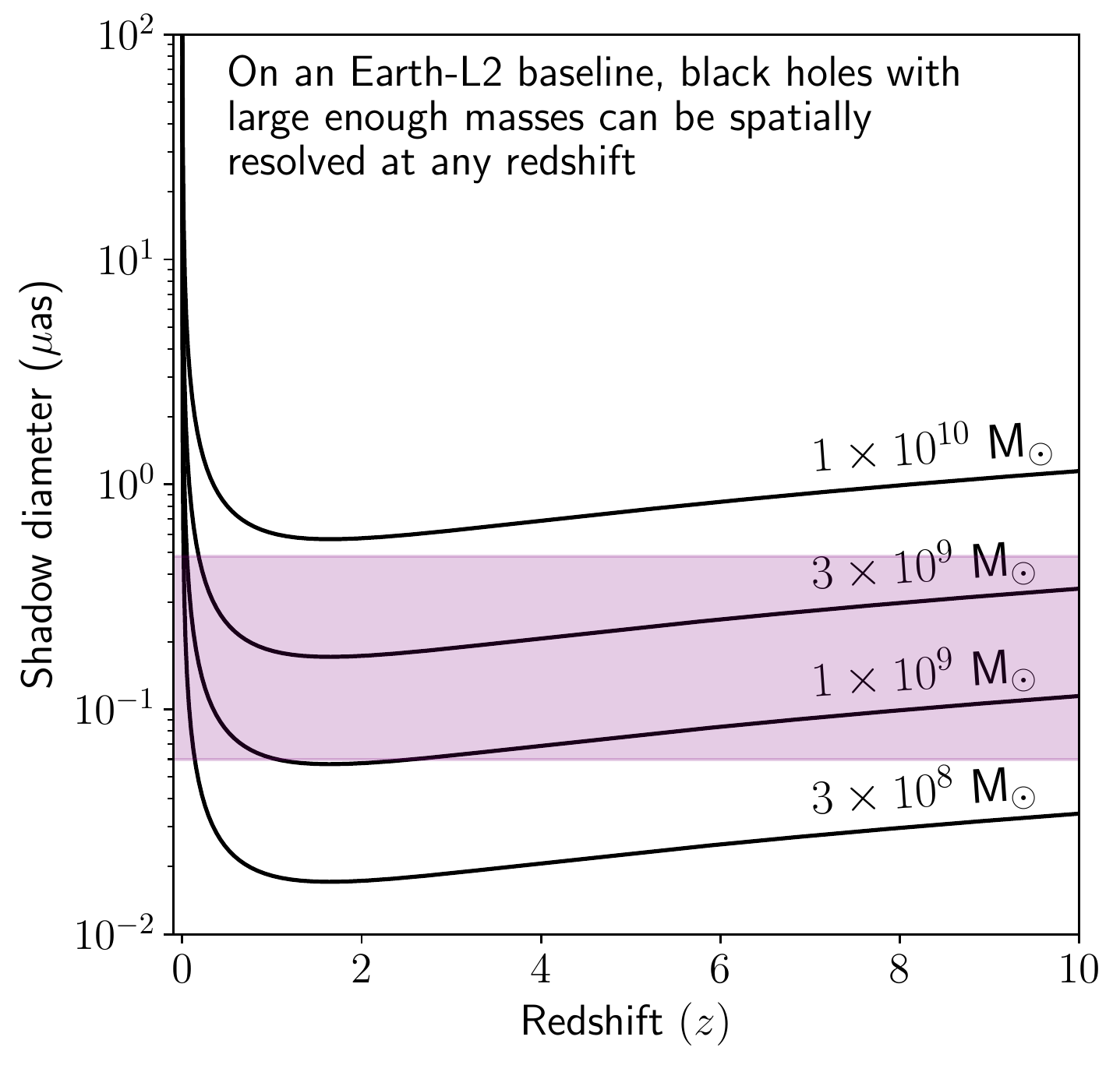}
\includegraphics[width=0.49\textwidth]{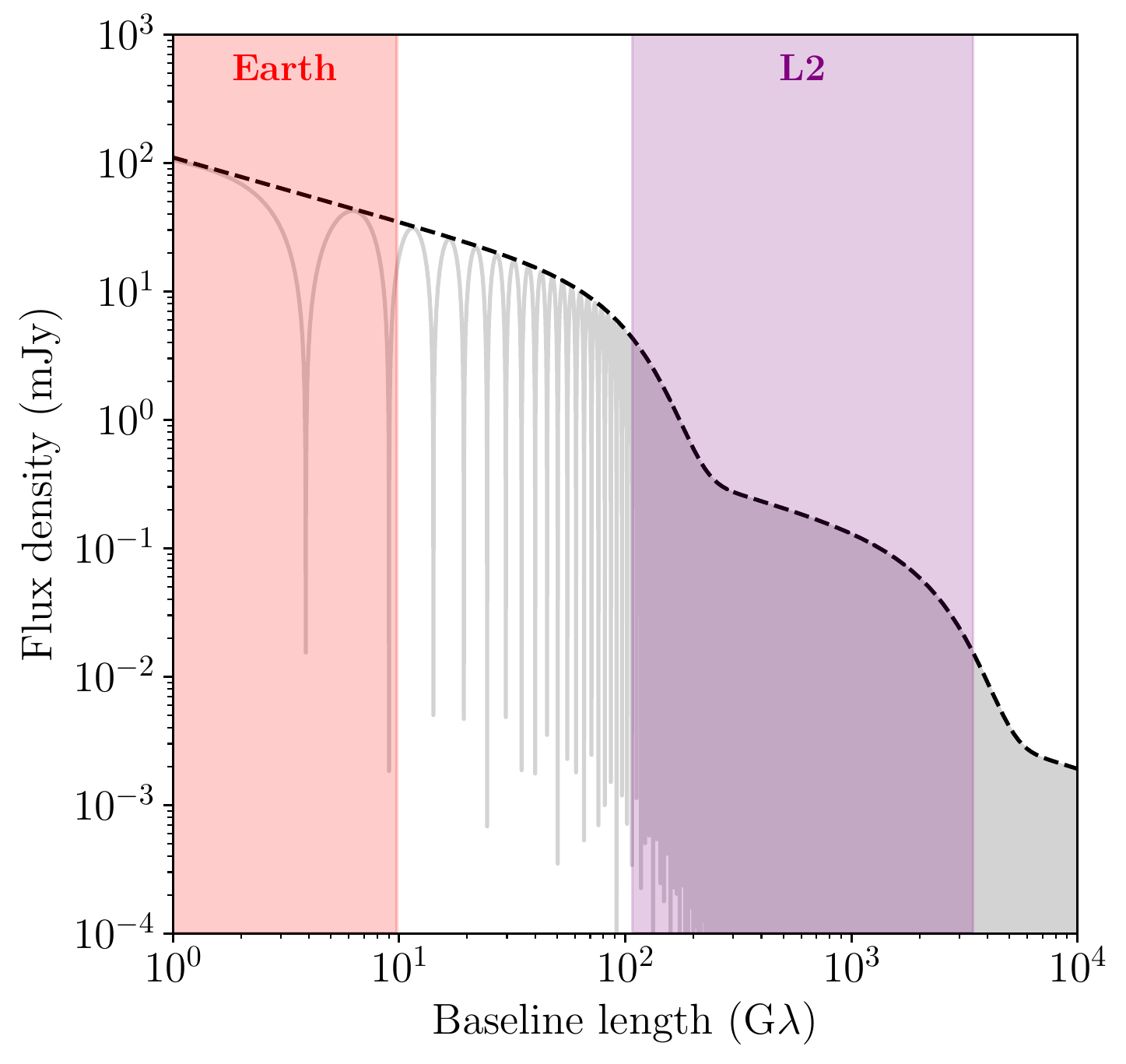}
\caption{\textit{Left}: Black hole shadow diameter vs. redshift for SMBHs of varying mass; the range of sizes accessible with a maximal Earth--L2 baseline operating at frequencies of 86--690\,GHz is shaded in purple. We can see that as shadows reach a minimum size at $z \approx 2$, SMBHs with masses greater than ${\sim}10^9$\,\msun are resolvable at all redshifts. \textit{Right}: Flux density as a function of baseline length is plotted in gray for a simple model of the \m87 photon ring structure \cite{Johnson_2019}, with the envelope shown as a dashed black line; note that the periodic structure evident at short baselines continues at longer baselines with the same period.  The purple shaded region shows the range of baselines accessible over the course of a year on an Earth--L2 baseline operating at frequencies of 86--690\,GHz; the long-baseline end of this range corresponds to a maximal baseline at 690 GHz, while the short-baseline end of the range corresponds to a minimal projected baseline (for \m87, the minimal projected baseline is a factor of $\sim$4 shorter than the maximal baseline) at 86 GHz.  For comparison, the range of baselines accessible from the ground at 230 GHz observing frequency is shown in red.  The Earth-L2 baseline coverage samples many more periods of the visibility structure than accessible from the ground, enabling correspondingly more precise measurements of the photon ring diameter.
\label{fig:angular_diameter_distance_and_photon_ring}}
\end{figure}

In our Universe, the angular diameter distance reaches a maximum value at a redshift of $z \approx 2$; sources of a given physical size thus have a minimum possible angular size, and if a source can be spatially resolved at $z \approx 2$ then it can be spatially resolved at any redshift.  The left panel of \autoref{fig:angular_diameter_distance_and_photon_ring} shows that on the extremely long baselines between Earth and L2, with fringe spacings $\theta \lesssim 0.1$\,\uas, this minimum angular size ensures that SMBHs with masses $M \gtrsim 10^9$\,\msun have shadow diameters that can be spatially resolved across cosmic history.  Measurements of the SMBH mass distribution as a function of redshift would shed light on SMBH-galaxy coevolution and inform cosmological models of structure formation, helping to understand how supermassive black holes formed in the early Universe (see, e.g., \citealt{Latif_2013}).

\subsection{Measuring SMBH spin and testing GR using photon ring substructure}

For the nearest SMBHs, such as M87, a baseline between Earth and L2 opens up the possibility of spatially resolving the substructure of the photon ring itself, \textbf{enabling new, stringent tests of GR}.  As detailed by \citet{Johnson_2019}, the self-similar nature of the photon ring is naturally decomposed by interferometers, with successive ``windings'' of photon orbits dominating the signal in discrete baseline intervals (see right panel of \autoref{fig:angular_diameter_distance_and_photon_ring}).  The period of the visibility signal along a particular orientation is a function of the ring diameter along that orientation.  The orbit of L2 around the Sun over the course of a year ensures that all orientations can in principle be sampled, enabling precise measurements of the photon ring size and shape.  The shape of the photon ring around a Kerr black hole is uniquely defined by its spin and inclination angle, with mass acting exclusively as an overall scaling factor. Measuring the ratio between shadow diameters at different orientations thus provides a sensitive probe of the SMBH spin as well as a way to test the validity of GR itself \citep{Johannsen_2010,Broderick_2014}.

\section{Technology overview}

\begin{figure}[t]
\centering
\includegraphics[width=0.50\textwidth]{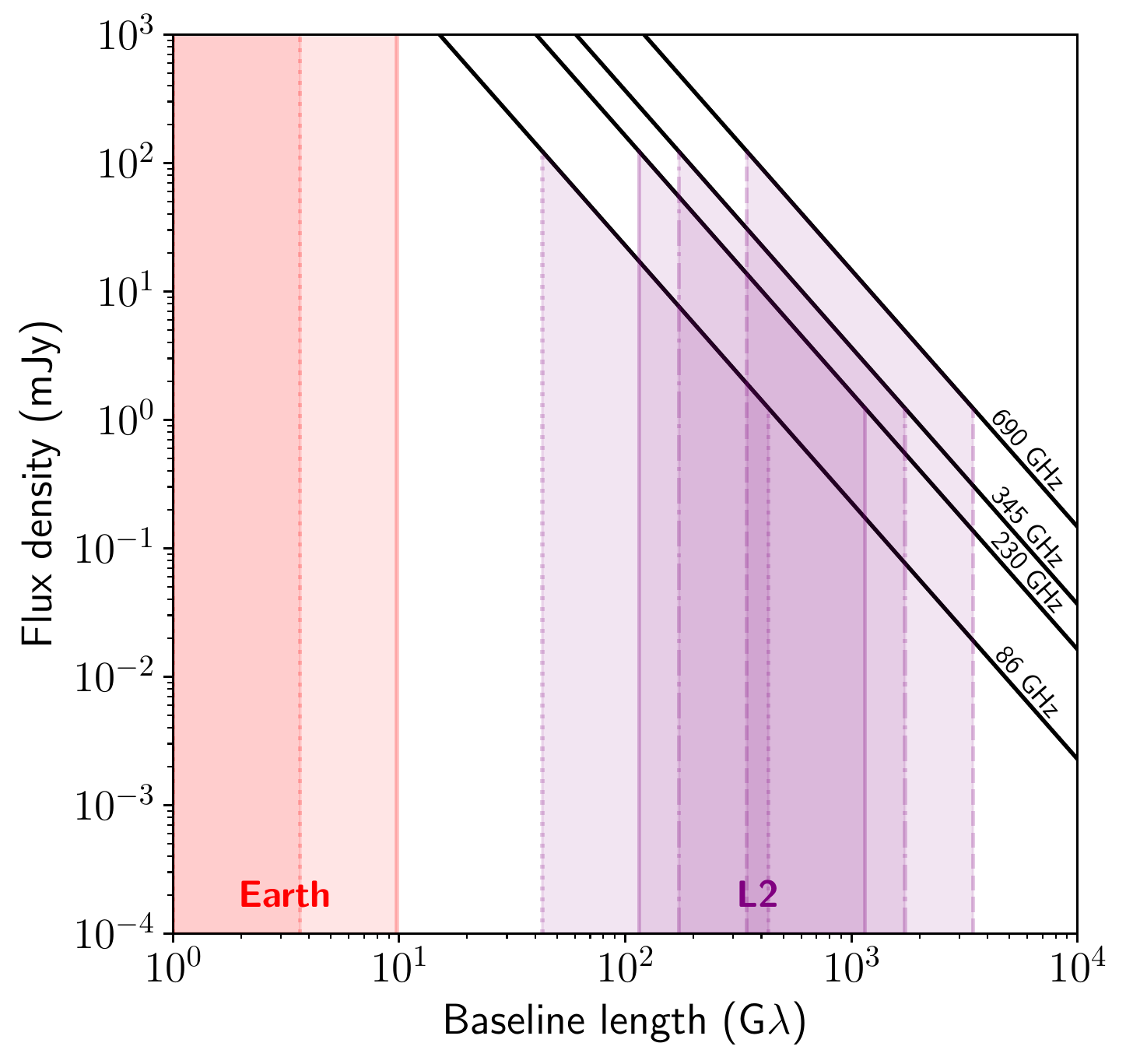}
\includegraphics[width=0.49\textwidth]{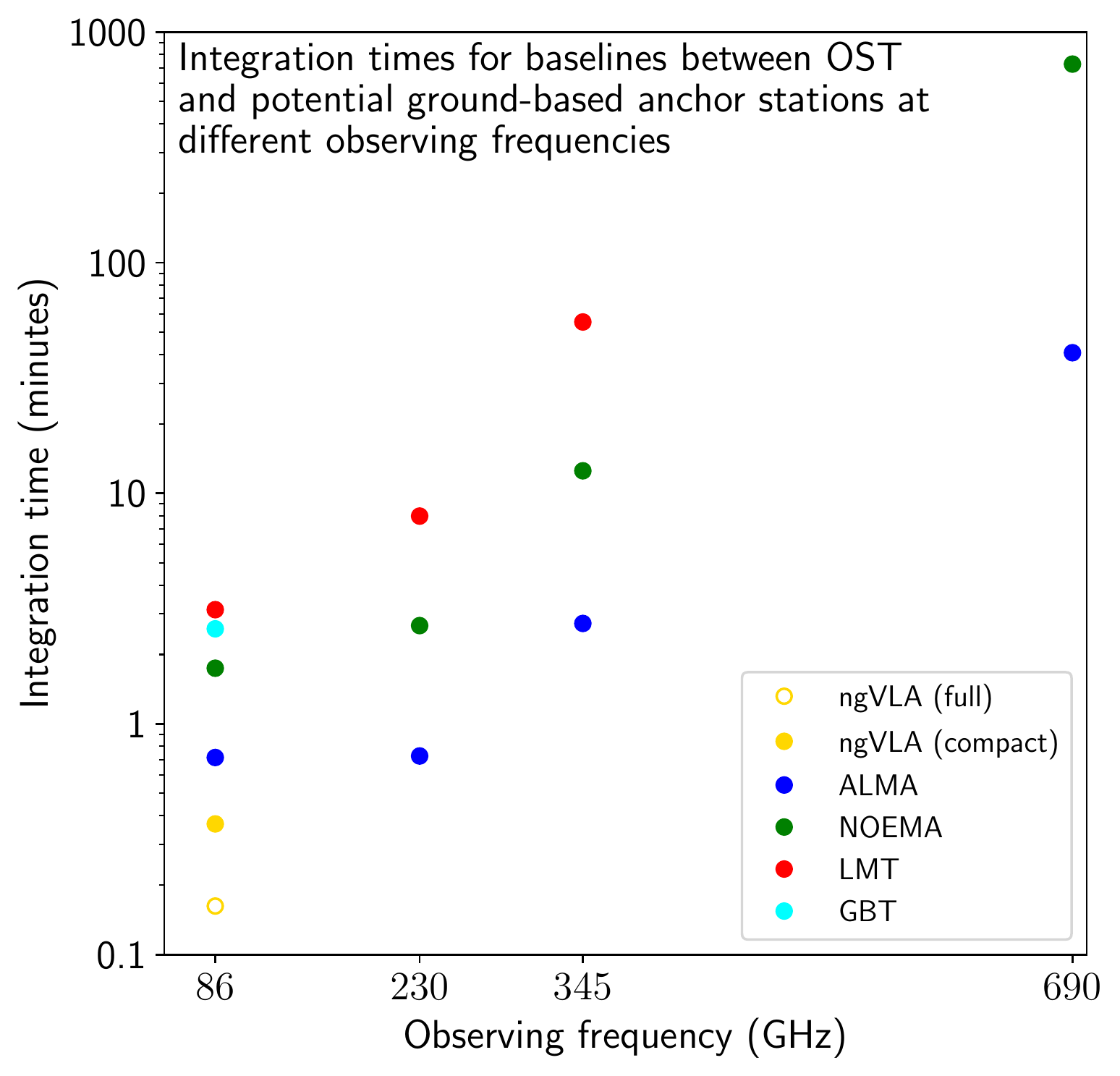}
\caption{\textit{Left}: For baselines to L2, synchrotron self-absorption and inverse-Compton scattering limit the brightness temperature to $T_b \lesssim 10^{12}$\,K \cite{Kellermann_1969}; the black lines mark this limit for each of the labeled observing frequencies.  Below each of these lines, the region shaded purple indicates physically allowed flux densities observable on Earth-L2 baselines ranging between 10--100\% of the maximum projected baseline length; overlapping regions indicate where sources could be observed with matched resolution at two or more frequencies (albeit at different times).  The red shaded regions are analogous to the purple ones, but for Earth-Earth baselines at 86 and 230 GHz only. \textit{Right}: Estimated integration times required to achieve a $5\sigma$ detection of a source with brightness temperature $T_b = 10^{12}$\,K on a maximal baseline between OST and various potential Earth-based anchor stations, as a function of observing frequency.  For the sensitivity estimates, we have assumed a total bandwidth (across all sidebands and polarizations) of 32 GHz for the 86 GHz observations, and 64 GHz for all other frequencies.  All ground stations are assumed to be observing at 45 degrees elevation, with zenith opacity of 0.05 (ALMA), 0.09 (NOEMA), 0.13 (LMT) at 230 GHz, and 0.05 (GBT, ngVLA) at 86 GHz. An aperture efficiency of 0.7 was assumed for all sites. Receiver temperatures are taken from station specifications or projections.
\label{fig:brightness_temp_and_sensitivity}}
\end{figure}

All extremely long baseline observations will face sensitivity challenges related to the physical brightness temperature limits of synchrotron radiation imposed by self-absorption and inverse-Compton scattering (\citealt{Kellermann_1969}; see left panel of \autoref{fig:brightness_temp_and_sensitivity}).  On a maximal\footnote{By ``maximal'' here, we mean the baseline length with no projection; only sources located at the ecliptic pole see exclusively maximal baselines, but every source in the sky sees a maximal baseline at some point in the orbit.} Earth--L2 baseline, no source is expected to have a flux density exceeding ${\sim}1$\,mJy. This strict sensitivity requirement drives the technology considerations.

The sensitivity of an interferometric baseline depends on: (1) the geometric mean of the system equivalent flux densities of the individual telescopes; (2) the averaged bandwidth; and (3) the coherent integration time. The first property allows telescopes such as \origins (at 5.9-meter diameter) to form sensitive baselines when paired with a large ground-based telescope (e.g., ALMA or ngVLA).  The second allows digital enhancements (e.g., wider recorded bandwidths) to offset limitations in telescope sensitivity. The third ties sensitivity to phase stability, which is limited by the atmosphere and reference frequency.

The following component recommendations are based on first-pass analyses and discussions, and are meant to give a general idea of the technology enhancements \origins would need to perform as an interferometric VLBI station.

\subsection{Receiver}

Atmospheric conditions above ground based stations limit the data collection frequencies to windows around 86 GHz, 230 GHz, 345 GHz, and 690 GHz. Band 1 of the HERO instrument already includes 690 GHz.  The remaining frequencies could be covered using two additional bands, one that spans the 86 GHz and 230 GHz windows, and another that covers 230 GHz through 345 GHz.  Simultaneous multifrequency observations will be useful as the ground-based array expands its capabilities to include multifrequency phase transfer.

\subsection{Data Processing/Storage/Downlink}

To achieve the sensitivities required for an Earth--L2 baseline, a combination of wide bandwidths and long integration times are needed. Though an improvement of one relaxes the requirement of the other, a first-pass operational configuration suggests that a total bandwidth of 64 GHz (16 GHz per sideband) for an integration time of two hours could achieve the requisite sensitivity on baselines to ALMA at any frequency (see right panel of \autoref{fig:brightness_temp_and_sensitivity}).  These requirements inform the data processing, storage/downlink, and timing reference components.

\subsubsection{Data processing}

Onboard data processing requires high-speed analog-to-digital converters (ADCs) and a high-performance Field Programmable Gate Array (FPGA). The FPGA quantizes and packetizes the digitized data for storage. Each of the four 8 GHz channels sampled at Nyquist requires a 16 Gbps ADC. The FPGA receives multiple streams and quantizes the data from the 4-bit sample to 2 bits, giving a total data rate of 256 Gbps. Current FPGA transceiver technology meets the speeds required, and both Xilinx (Ultrascale Kintex) and Microsemi (RTG-4) are currently working to make these high-speed interfaces available in space-qualified components.

\subsubsection{Storage/Downlink}

With an observation time of 6 hours, a 1/3 duty cycle and a rate of 256 Gbps, a total of approximately 230 TB of data will be captured. A tradeoff between onboard storage and downlink speed will need to be conducted to determine the most optimal design. Storage continues to increase in both capacity and speed, with 1TB currently available in a small (22 mm $\times$ 80 mm) package. Downlink capabilities, pushed by the telecommunications industry, are also getting faster with laser communication speeds currently at 200 Gbps from LEO to small ground-based receivers \citep{RobinsonB.S2018TIDT}.

\subsection{Timing reference}

To achieve phase stability on a single baseline for integration times of $\sim$hours requires either that both stations are individually equipped with extremely stable timing references or that they share a common reference. Individual clocks would need to have Allan deviations better than ${\sim}5 \times 10^{-18}$ over the integration time, which is roughly three orders of magnitude more stable than current ground-based technology and likely unachievable on a 1--2 decade timescale.  Instead, a shared reference would relax the more stringent stability requirement almost entirely and will likely prove to be the more feasible option.  A shared ground-space timing reference has been demonstrated at lower observing frequencies by RadioAstron \citep{Kardashev_2013}, though additional work would be required to extend such capabilities to higher observing frequencies and to a station at L2.

\subsection{Positional/Velocity Accuracy}

Antenna position and velocity (and possibly higher order derivatives) must be known in order to coherently average the correlated signal across finite bandwidth and over time. Initial searches can be conducted with wide search windows in the associated delay, delay-rate, and acceleration parameters, with residual values being used to refine the orbit determination, as is currently done for RadioAstron \citep{Zakhvatkin_2018}. To create a baseline of position, velocity, and acceleration requirements, we take the RadioAstron specifications and increase the requirement by a factor of 10 for velocity and acceleration:

\begin{itemize}
    \item Position error less than 600 m
    \item Velocity error less than 2 mm\,s$^{-1}$
    \item Acceleration error less than $10^{-9}$ m\,s$^{-2}$
\end{itemize}

\noindent These requirements can be potentially relaxed with improvements in computational delay and rate searching routines at the correlation stage.

\section{Summary}
Extremely long baseline interferometry with \origins would create new and exciting possibilities for the study of black holes and general relativity under the most extreme conditions.  Moreover, the technology required to enable this possibility appears either feasible today, or within reach during the time \origins would be built.  Given the immense scientific potential, it is our hope that the Decadal Review will seize this opportunity and recommend that this capability be considered for inclusion in \origins.

\pagebreak

\printbibliography

\end{document}